\documentclass[twocolumn,journal]{IEEEtran}
\usepackage[T1]{fontenc}
\usepackage{array}
\usepackage{textcomp}
\usepackage{amsmath}
\usepackage{graphicx}
\usepackage[unicode=true,
 bookmarks=true,bookmarksnumbered=true,bookmarksopen=true,bookmarksopenlevel=1,
 breaklinks=false,pdfborder={0 0 0},backref=false,colorlinks=false]
 {hyperref}
\hypersetup{pdftitle={Your Title},
 pdfauthor={Your Name},
 pdfpagelayout=OneColumn, pdfnewwindow=true, pdfstartview=XYZ, plainpages=false}
\usepackage{breakurl}

\makeatletter

\providecommand{\tabularnewline}{\\}

 \let\oldforeign@language\foreign@language
 \DeclareRobustCommand{\foreign@language}[1]{%
   \lowercase{\oldforeign@language{#1}}}

\usepackage[caption=false,font=footnotesize]{subfig}

\makeatother

\begin{document}

\title{Towards Implementing Multi-Channels, Ring-Oscillator-Based, Vernier
Time-to-Digital Converter in FPGAs: Key Design Points and Construction
Method}

\author{Ke~Cui, Xiangyu\,Li, Zongkai~Liu and~Rihong~Zhu\thanks{This work was supported by the Fundamental Research Funds for the
Central Universities under Grants 30916014112-019 and 30916011349.}\thanks{Ke~Cui, Zongkai~Liu and Rihong~Zhu are are with the MIIT Key Laboratory
of Advanced Solid Laser, Nanjing University of Science and Technology,
Nanjing, Jiangsu, China, and also with the Advanced Launching Co-innovation
Center, Nanjing University of Science and Technology, Nanjing, Jiangsu,
China (e-mail: \protect\href{http://njustcuik@njust.edu.cn}{njustcuik@njust.edu.cn}).}\thanks{Xiangyu~Li is with the School of Computer Science and Engineering,
Nanjing University of Science and Technology, Nanjing, Jiangsu, China.}}

\markboth{}{}

\IEEEpubid{}
\maketitle
\begin{abstract}
For TOF positron emission tomography (TOF PET) detectors, time-to-digital
converters (TDCs) are essential to resolve the coincidence time of
the photon pairs. Recently, an efficient TDC structure called ring-oscillator-based
(RO-based) Vernier TDC using carry chains was reported by our team.
The method is very promising due to its low linearity error and low
resource cost. However, the implementation complexity is rather high
especially when moving to multi-channels TDC designs, since this method
calls for a manual intervention to the initial fitting results of
the compilation software. In this paper, we elaborate the key points
toward implementing high performance multi-channels TDCs of this kind
while keeping the least implementation complexity. Furthermore, we
propose an efficient fine time interpolator construction method called
the period difference recording which only needs at most 31 adjustment
trials to obtain a targeted TDC resolution. To validate the techniques
proposed in this paper, we built a 32-channels TDC on a Stratix III
FPGA chip and fully evaluated its performance. Code density tests
show that the obtained resolution results lie in the range of (23
ps \textasciitilde{} 37 ps), the differential nonlinearity (DNL) results
lie in the range of (-0.4 LSB \textasciitilde{} 0.4 LSB) and the integral
nonlinearity (INL) results lie in the range of (-0.7 LSB \textasciitilde{}
0.7 LSB) for each of the 32 TDC channels. This paper greatly eases
the designing difficulty of the carry chain RO-based TDCs and can
significantly propel their development in practical use. \end{abstract}


\IEEEpeerreviewmaketitle{}

\section{Introduction}

\IEEEPARstart{T}{ime}-of-flight (TOF) PET consists of very fast
detectors which utilize multi-channel time-to-digital converter (TDC)
modules to resolve the coincidence time of the photon pairs. This
helps to improve the tomographic reconstruction quality while simultaneously
reducing radiation doses and/or scan times \cite{Dan2009Impact}.
The performance of the overall PET system is directly related to the
precision of the used TDCs. In this paper, we will focus on the designing
of highly accurate multi-channel TDCs while keeping as least implementation
complexity as possible. 

Most present TDC structures adopt a two-step time measurement technique
\cite{Wu2010}-\nocite{1610982}\nocite{LZhao2013}\cite{YonggangW2016-2}.
In this method, the first step uses a coarse counter running at system
clock rate (usually corresponding to a period of several nanoseconds)
to record the elapsed coarse time to guarantee large dynamic range.
The second step adopts a fine time interpolator with subnanosecond
resolution to accurately record the time bin locating in the specific
system clock cycle at which the coarse time counter is latched to
guarantee high precision. The mostly used fine time interpolator techniques
include: tapped delay line (TDL) \cite{1610982}\nocite{LZhao2013}\nocite{YonggangW2016-2}\nocite{JWu2008}\nocite{JWu2003}\nocite{5941022}\nocite{7057685}\nocite{QiShen2015}\nocite{WPan2014}-\cite{MFishburn2013},
pulse shrinking delay line \cite{SzpletKlepacki2010} and Vernier
delay line \cite{552156}\nocite{MarkovicTisaVillaEtAl2013}-\cite{JYu2010}.
Generally, there are two platforms to implement TDCs: application
specific integrated circuits (ASICs) and field programmable gate arrays
(FPGAs). ASIC-based TDCs have strong design flexibility and can utilize
some very beneficial analog circuits such as delay locked loops (DLLs)
contributing to an excellent delay line. However, their cost is especially
high when the production volume is low and the development period
is rather long. FPGA-based TDCs have much less design freedom which
are constrained in the digital design space. However, the reconfigurability
of FPGAs makes the design much less expensive and can be adjusted
to meet new requirements quickly.

\begin{figure*}[tbh]
\centering\includegraphics[width=0.9\textwidth]{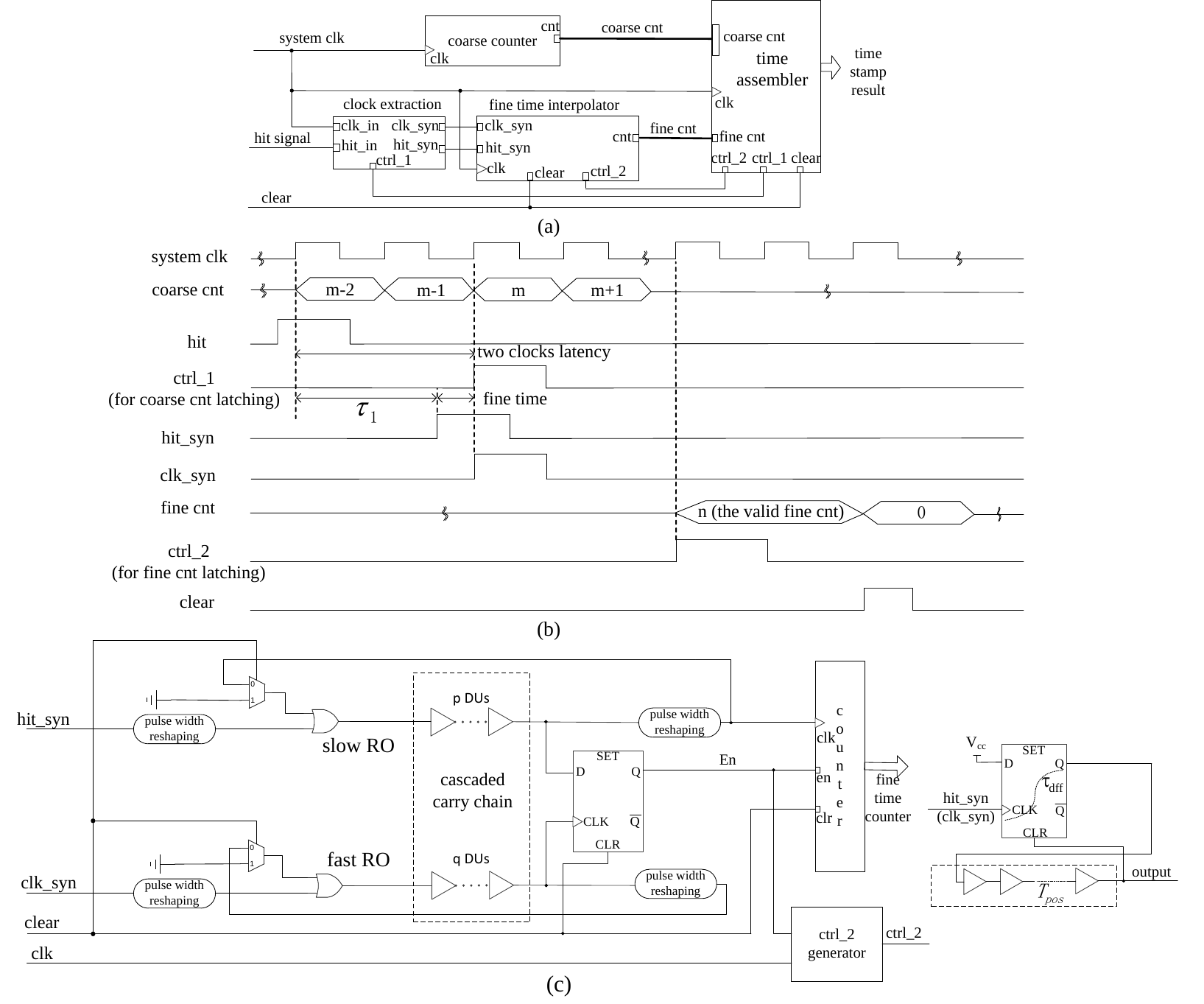}

\caption{Carry chain RO-based TDC. (a) overall structure; (b) timing diagram;
(c) structure of the RO-based fine time interpolator.}
\end{figure*}

J. Wu proposed the carry chain based structure as an efficient TDL
interpolator which is important to the development of the FPGA-based
TDCs \cite{JWu2003}. The carry chains which widely exist in modern
FPGAs are specially provided by their vendors to fulfill fast algorithm
functions such as fast addition or comparison. The delay time of a
basic carry chain cell is very small such that it is reasonably conceived
as the most ideal tool to fulfill the fine time interpolation task
on FPGA chips. That is the reason why the carry chain based TDL TDCs
have gained extensive studies in recent years. However, existence
of ultra-wide bin which is physically determined during chip fabrication
process limits the precision of such TDCs significantly. J. Wu proposed
the wave-union A and B methods to effectively subdivide the ultra-wide
bin by multiple measurements along a single carry chain and improved
the precision beyond its cell delay \cite{JWu2008}. The wave-union
methods are then widely adopted by many later emerging FPGA-based
TDC designs \cite{5941022}, \cite{WPan2014}. Another drawback of
such TDCs is the large differential nonlinearity (DNL) and integral
nonlinearity (INL) problem caused by uneven bin granularity of the
used carry chains. One possible technique to mitigate this problem
is to use the bin-by-bin calibration techniques \cite{Wu2010}. However,
this incurs large memory and logic resource cost.

Recently, we proposed a new ring-oscillator-based (RO-based) TDC structure
by organizing the carry chains in a Vernier loop style \cite{Cui2017}.
A specific construction method to set up two ROs with very little
period difference was fully illustrated. This method opened up a new
way to utilize the carry chains to build the fine time interpolator
which led to much reduced DNL and INL. In this paper, we report a
32-channels TDC realized on a single FPGA chip by further exploiting
the method proposed in \cite{Cui2017}. One main challenge on the
implementation complexity emerges when moving to multi-channels TDC
designs, since the efforts which should be paid linearly increase
with the channel number. This is especially true because the design
flow requires an exhaustive manual intervention to the initial fitting
results of the compilation software to obtain the two ROs with a targeted
period difference. This paper proposes a new construction method for
building the two ROs with definitely less than 31 trials per TDC channel
which greatly reduces the design complexity. In summary, the main
contributions of this paper contain: key points to obtain high performance
TDC by utilizing the carry chain RO-based method; a new and highly
efficient construction method called the period difference recording
(PDR) to build the ROs; multi-channel ability and scalability by utilizing
the carry chain RO-based method in a single FPGA chip.

\section{Multi-channel TDC Design}

\subsection{Carry Chain RO-Based TDC Structure}

The basic carry chain RO-based Vernier TDC structure is depicted in
Fig.1. It uses two steps to measure a time interval including a coarse
and a fine time measurement steps (Fig.1(a)). The corresponding timing
diagram is shown in Fig.1(b). The coarse counter running at the system
clock rate is adopted to record the coarse time. The clock extraction
module is designed to find the closest clock signal in time after
the hit signal and extract the delayed hit and clock signals pair
to the fine time interpolator module to measure the fine time interval
between them. The working principle and circuit implementation of
the clock extraction module can be found in Section II-B. Two signals
labeled ctrl\_1 and ctrl\_2 are generated to denote the proper timing
for the time assembler module to latch the coarse and fine counter
values correctly and combine them together to produce the final timestamp.
Fig.1(c) shows the detailed structure of the RO-based Vernier fine
time interpolater which connects the last cell of the carry chain
back to its first cell. The two ROs are composed of different numbers
of carry chain cells (or delay units - DUs) and hold different oscillation
periods. The DU works as basic delay unit and a complete Vernier delay
line can contain an even or odd number of DUs. The period difference
between the two ROs determines the resolution of the TDC. Additionally,
each RO contains a pulse width reshaping module to maintain the stability
of the positive duration of the oscillation signal propagating along
it. Its circuit implementation is shown in the rightmost part of Fig.1(c).
According to the timing diagram (Fig.1(b)), the leading signal of
a fine time interval event (the hit\_syn signal in Fig.1) is fed to
the slow RO while the lagging one (the clk\_syn signal in Fig.1) to
the fast RO. The fine timestamp is obtained by reading out the fine
time counter which records the oscillation number at the moment that
the lagging signal catches up the leading signal. Obviously, the DNL
of such TDCs avoids the bad influence of the uneven bin granularity,
since the resolution is determined by the physical length difference
of the two ROs but not the bin widths of the used carry chains as
in the TDL form. Much reduced DNL has been observed in \cite{Cui2017}.
However, since the ROs are not compensated and stabilized during the
time measurement, the oscillation number cannot be set too large to
assure small precision RMS which means that tradeoff between the resolution
and the precision should be carefully considered and made by the designers.

\subsection{Key Design Points}

There are two key points for the designers to build the RO-based Vernier
TDC: the clock extraction module and the fine time interpolator module.

\subsubsection{key design point for the clock extraction module}

\begin{figure}[tbh]
\centering\includegraphics[width=0.95\columnwidth]{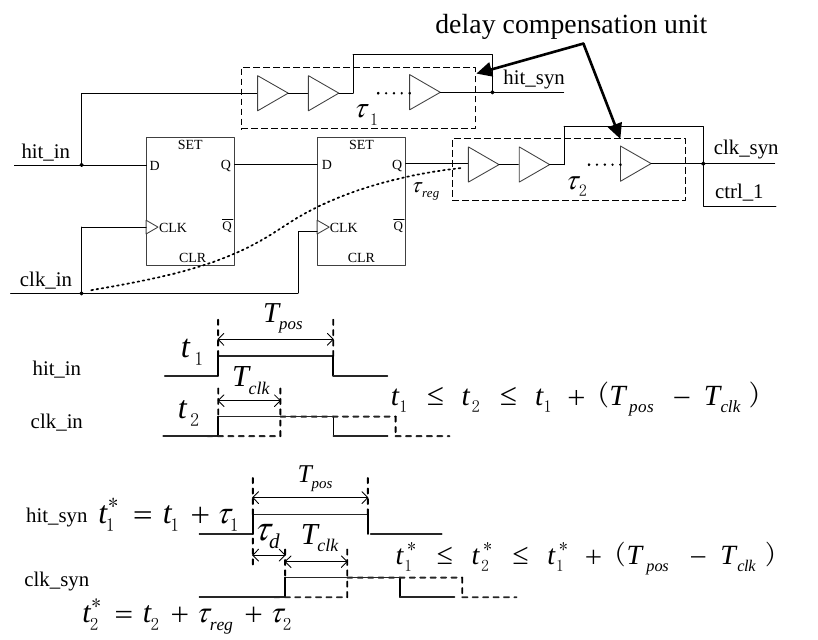}

\caption{Circuit implementation and timing diagram of the clock extraction
module. Here $t_{1}$ and $t_{2}$ are the arrival time of the hit\_in
and clk\_in signals respectively, while $t_{1}^{*}$ and $t_{2}^{*}$
are the corresponding output time after passing the clock extraction
module.}
\end{figure}

\begin{figure}[tbh]
\centering\includegraphics[width=0.95\columnwidth]{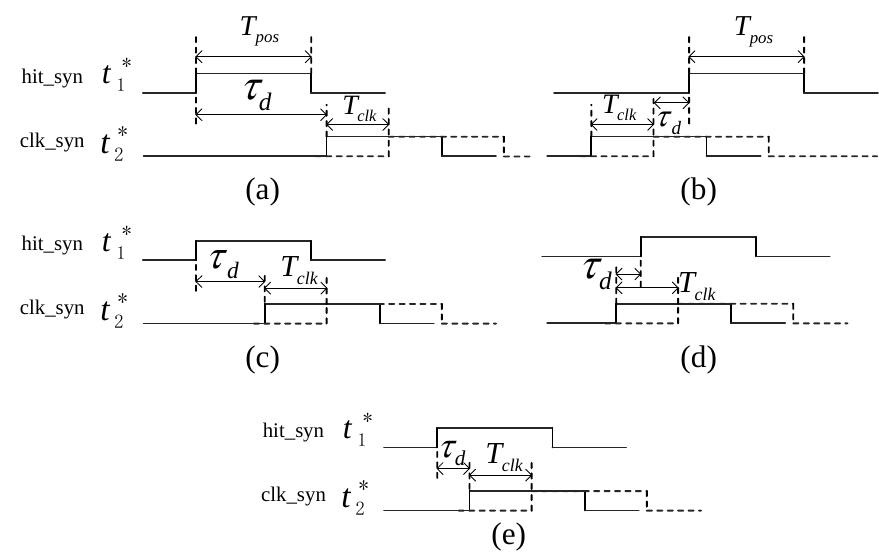}

\caption{Relative phase relationship between the hit\_syn and clk\_syn signals.}
\end{figure}

The basic task for the clock extraction module is to locate the clock
signal which is nearest in time after the hit signal and then extract
it out. The outputted leading and lagging signals are termed as hit\_syn
and clk\_syn correspondingly and the fine time interval between them
will be measured by the following fine time interpolator module. The
circuit implementation of the clock extraction module is depicted
in Fig.2. It can be seen that an undesired additional delay $\tau_{d}=\tau_{reg}+\tau_{2}-\tau_{1}$
is added to the original fine time interval and causes a minimal oscillation
number $n_{0}=\frac{\tau_{d}}{\textrm{LSB}}$, where $\tau_{reg}$
represents the delay introduced by the sampling D-type flip-flops,
$\tau_{1}$ and $\tau_{2}$ represents the adjustable delay introduced
by the delay compensation units for the hit\_in and clk\_in signals
respectively, and LSB represents the resolution of the TDC. The existence
of $n_{0}$ deteriorates the TDC performance, since the precision
RMS increases about proportionally to the square root of the overall
oscillation number \cite{AAAbidi2006}. The timing diagram in Fig.2
shows that the following formula should be satisfied to guarantee
the requirement that the renewed arrival time of the clk\_syn signal
should not exceed the positive duration of the hit\_syn signal for
correct operating of the sampling D-type flip-flop in the fine time
interpolator:

\begin{equation}
0\leq\tau_{d}\leq(T_{pos}-T_{clk})\label{eq:clk_timing}
\end{equation}

where $T_{pos}$ represents the positive duration of the hit\_syn
signal and $T_{clk}$ represents the period of the system clock. The
compensation delays $\tau_{1}$ and $\tau_{2}$ are intentionally
added to try to make $\tau_{d}$ as small as possible leading to the
least $n_{0}$ to gain the best precision performance. The compensation
unit is composed of 32 cascaded look-up table (LUT) implemented NOT
gates. According to our experimental experience, this gate amount
is adequate to find a good enough parameter set $(\tau_{1},\tau_{2})$.
The actually used gates number for each signal is manually adjusted
and determined by using the resource editor tool provided by the FPGA
manufacturers (for example the engineering change orders - ECO tool
by Altera and the FPGA editor tool by Xilinx). The adjustment criteria
is to make $\tau_{d}$ locating in the range constrained by equation
(\ref{eq:clk_timing}) and as close to zero as possible. However,
this task is difficult to be fulfilled directly since all the timing
parameters $\tau_{1}$, $\tau_{2}$, $\tau_{reg}$ and $T_{pos}$
are very hard to be known exactly. To combat the difficulty, we propose
to infer the actual $\tau_{d}$ value by observing the distribution
of the outputted fine time counter values $n$ captured from the fine
time interpolator module. The cases of the relative phase between
the hit\_syn and clk\_syn signals are depicted in Fig.3 and the corresponding
distributions of $n$ are summarized in Table I, where the parameter
$n_{m}$ represents the maximal fine time counter value.

\begin{table}[tbh]
\caption{Distributions of the fine time counter values $n$ corresponding to
different cases in Fig.3}

\centering

\begin{tabular}{|>{\centering}m{0.1\columnwidth}|>{\centering}m{0.5\columnwidth}|>{\centering}m{0.25\columnwidth}|}
\hline 
case number & description & distribution of $n$\tabularnewline
\hline 
\hline 
(a) & all the occurring time points of the clk\_syn are late to the positive
duration of the hit\_syn & \{0\}\tabularnewline
\hline 
(b) & all the occurring time points of the clk\_syn are prior to the positive
duration of the hit\_syn & \{0\}\tabularnewline
\hline 
(c) & part of the occurring time points of the clk\_syn are late to the
positive duration of the hit\_syn & \{0\} $\cup$ \{$(n_{0},\thinspace n_{m})$\}\tabularnewline
\hline 
(d) & part of the occurring time points of the clk\_syn are prior to the
positive duration of the hit\_syn & \{$(0,\thinspace n_{m})$\}\tabularnewline
\hline 
(e) & all the relative phase relations are in the expect range satisfying
equation (\ref{eq:clk_timing}) & \{$(n_{0},\thinspace n_{m})$\}\tabularnewline
\hline 
\end{tabular}
\end{table}

According to the cases listed in Table I, the following steps are
applied to efficiently find an optimal set $(\tau_{1},\tau_{2})$
for each TDC channel:

\begin{figure*}[tbh]
\centering\includegraphics[width=0.95\textwidth]{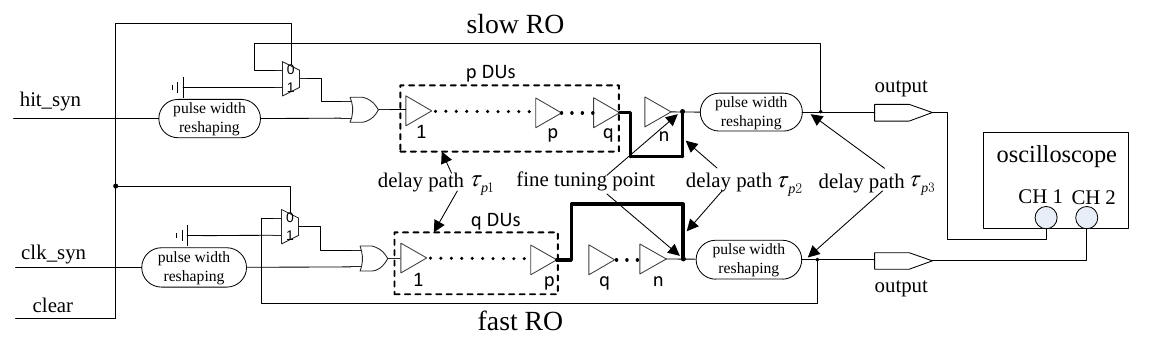}

\caption{Delay paths along the RO.}
\end{figure*}

\begin{enumerate}
\item According to the collected distributions of $n$, classify which case
the present parameter set $(\tau_{1},\tau_{2})$ belongs to. If it
accords with case (c), goto step 2; if it accords with case (d), goto
step 3; if it accords with case (e), goto step 4; otherwise modify
the number of cascaded NOT gates in either of the two delay compensation
unit until one of the cases (c)\textasciitilde{}(e) appears. 
\item Iteratively shorten the gates number of the delay compensation unit
in the clk\_syn signal path until case (e) appears and then goto step
4.
\item Iteratively shorten the gates number of the delay compensation unit
in the hit\_syn signal path until case (e) appears and then goto step
4.
\item Decrease $n_{0}$ by interatively shortening the gates number of the
delay compensation unit in the clk\_syn signal path until the minimal
achievable positive $n_{0}$ is found. 
\end{enumerate}
The shortened gates number in steps 2 \textasciitilde{} 3 is usually
set larger than 1 to boost the finding process while the number in
step 4 is set just as 1 to search the optimal $n_{0}$. The above
mentioned manual adjustment for the clock extraction module is very
useful, since as the targeted TDC channels number increases, the initial
fitting results have more risk to lay out of case (e) and lead to
operation failure. Even the automatic fitting results initially accords
with case (e), it is still very beneficial to apply step 4 to find
the optimal $n_{0}$.

\subsubsection{key design point for the fine time interpolator module}

The fine time interpolator module contains two structure-symmetric
ROs built from the carry chains. The structural similarity is obtained
by utilizing the partition based two-step construction method proposed
in \cite{Cui2017}. The complete oscillation period of the RO is composed
of three parts: $\tau_{p1}$ caused by the carry chain (the path encompassed
by the dotted rectangle), $\tau_{p2}$ caused by the connection path
between the end of the carry chain and the pulse reshaping module
(the bold line), and $\tau_{p3}$ caused by all the remaining logic
units and paths in the RO as shown in Fig.4. 

It is clear that $\tau_{p3}$ keeps constant once the RO is set up.
In our previous work, $\tau_{p2}$ is also assumed to be unchanged
after manual intervention at the fine tuning point, so an iterative
adjustment process of assigning different DU number combination sets
to alter $\tau_{p1}$ for the two ROs is adopted. The adjustment task
is performed as follows: cut off the connection at the fine tuning
point whose oscillation period is longer; shorten the length of the
carry chain by one DU; finally reconnect the new shorter carry chain
to the corresponding pulse width reshaping module. The oscillation
periods are observed on an external oscilloscope by introducing the
oscillation signals out of the FPGA chip. This adjustment principle
restricts that the DU number combination set assigning direction can
only be conducted forward to the front end of the carry chain, which
may incur the missing of many potential DU number combination sets,
since $\tau_{p2}$ becomes actually uncertain after each time of adjustment.
This arises from the possibility that when a RO needs to reduce its
oscillation period, the DU number may actually require adding 1 instead
of subtracting 1, if $\tau_{p2}$ decreases so dramatically that the
entire oscillation period decreases even with the larger DU number.
In our example design, the overall length of a complete carry chain
is 32 and this theoretically gives $32\times32=1024$ possible DU
number combination sets if both $\tau_{p1}$ and $\tau_{p2}$ are
viewed changeable. The release of the adjustment constraint generates
huge DU number combination set space and gives flexible design freedom.
This point is also important in multi-channels TDC designs since the
more DU number combination candidates can be used, the less design
failure may be encountered when using such a short carry chain (totally
32 DUs). If a design failure happens for a TDC channel, the designer
has to re-allocate a new physic region on the FPGA chip and re-construct
this bad channel which will greatly increase the design complexity.
Although extending the length of the used carry chain is also feasible
to improve the design success rate, it will cause much larger resource
cost which is especially true in multi-channels TDC designs.

\subsection{Period difference recording method for fine time interpolator construction}

In this section we propose the PDR method, by using which every possible
DU number combination set can be covered with very few total adjustment
trials. To clarify the PDR method, we define the oscillation period
of the fast RO as $\tau_{f,\thinspace i}$ ($i=32,\thinspace31,\thinspace...,\thinspace1$),
when the $i$-th DU number of the fast RO is connected to the fine
tuning point. Similarly we define the oscillation period of the slow
RO as $\tau_{s,\thinspace j}$ ($j=32,\thinspace31,\thinspace...,\thinspace1$),
when the $j$-th DU number of the slow RO is connected to the fine
tuning point. Additionally we define the oscillation period difference
between the fast and slow ROs as $\varDelta\tau_{i,\thinspace j}=\tau_{s,\thinspace j}-\tau_{f,\thinspace i}$
corresponding to the DU number combination set ($i,\thinspace j$).
By using the above definitions, we illustrate the PDR design flow
as follows:
\begin{enumerate}
\item Test and record the result of $\varDelta\tau_{32,\thinspace32}$,
and goto step 2.
\item Fix $i=32$, enumerate $j=31,\thinspace30,\thinspace...,\thinspace1$,
test and record the results of $\varDelta\tau_{32,\thinspace j}$,
and goto step 3.
\item Fix $j=32$, and initialize $i=31$ if this is the first time entering
this step. Test and record the result of $\varDelta\tau_{i,\thinspace32}$,
goto step 4.
\item For the current targeted DU number combination set $(i,\thinspace j)$
($j=31,\thinspace30,\thinspace...,\thinspace1$), compute $\varDelta\tau_{i,\thinspace j}=\varDelta\tau_{32,\thinspace j}+\varDelta\tau_{i,\thinspace32}-\varDelta\tau_{32,\thinspace32}$.
If any $\varDelta\tau_{i,\thinspace j}$ lying in the targeted resolution
range exists, output the combination set $(i,\thinspace j)$ and stop
the iteration with success, otherwise make $i=i-1$ and goto step
3. If no satisfying $\varDelta\tau_{i,\thinspace j}$ can be found
even with $i=1$, stop the iteration with failure. 
\end{enumerate}
The PDR method only needs to record at most 63 results but can cover
as much as 1024 different DU number combination sets, which greatly
reduces the design complexity. It originates from the following identical
equation:

\begin{equation}
\begin{array}{c}
\varDelta\tau_{i,\thinspace j}=\tau_{s,\thinspace j}-\tau_{f,\thinspace i}\\
=\tau_{s,\thinspace j}-\tau_{f,\thinspace32}+\tau_{s,\thinspace32}-\tau_{f,\thinspace i}+\tau_{f,\thinspace32}-\tau_{s,\thinspace32}\\
=\varDelta\tau_{32,\thinspace j}+\varDelta\tau_{i,\thinspace32}-\varDelta\tau_{32,\thinspace32}
\end{array}\label{eq:RDR}
\end{equation}

In practical use, the period difference between the two ROs is obtained
by observing an external oscilloscope to collect the oscillation number
$k$, the initial period difference $\varDelta\tau_{ini}$ and the
final period difference $\varDelta\tau_{fnl}$ for an arbitary DU
number combination set $(i,\thinspace j)$ ($i,\thinspace j=32,\thinspace31,\thinspace..,1$),
and then the period difference is calculated as $\varDelta\tau_{i,\thinspace j}=\frac{\varDelta\tau_{fnl}-\varDelta\tau_{ini}}{k}$.
For example, Fig.5 shows a real waveform captured during our design
process by a 2.5 Gs/s Tektronix oscilloscope (series number: DPO 3032),
of which the channel 1 represents the slow RO while the channel 2
represents the fast RO. It can be seen that Fig.5(a) shows the entire
oscillations waveform giving $k=25$, Fig.5(b) shows the locally enlarged
waveform of the first two oscillations giving $\varDelta\tau_{ini}\approx300$
ps, and Fig.5(c) shows the locally enlarged waveform of the last two
oscillations giving $\varDelta\tau_{fnl}\approx1000$ ps, so $\varDelta\tau_{i,j}$
can be estimated as $\frac{1000\thinspace\textrm{ps}-300\thinspace\textrm{ps}}{25}=28$
ps. Fig.5 also shows that the realized ROs have an approximately 200
MHz frequency (corresponding to an about 5 ns period). 

\begin{figure}[tbh]
\centering\includegraphics[width=0.95\columnwidth]{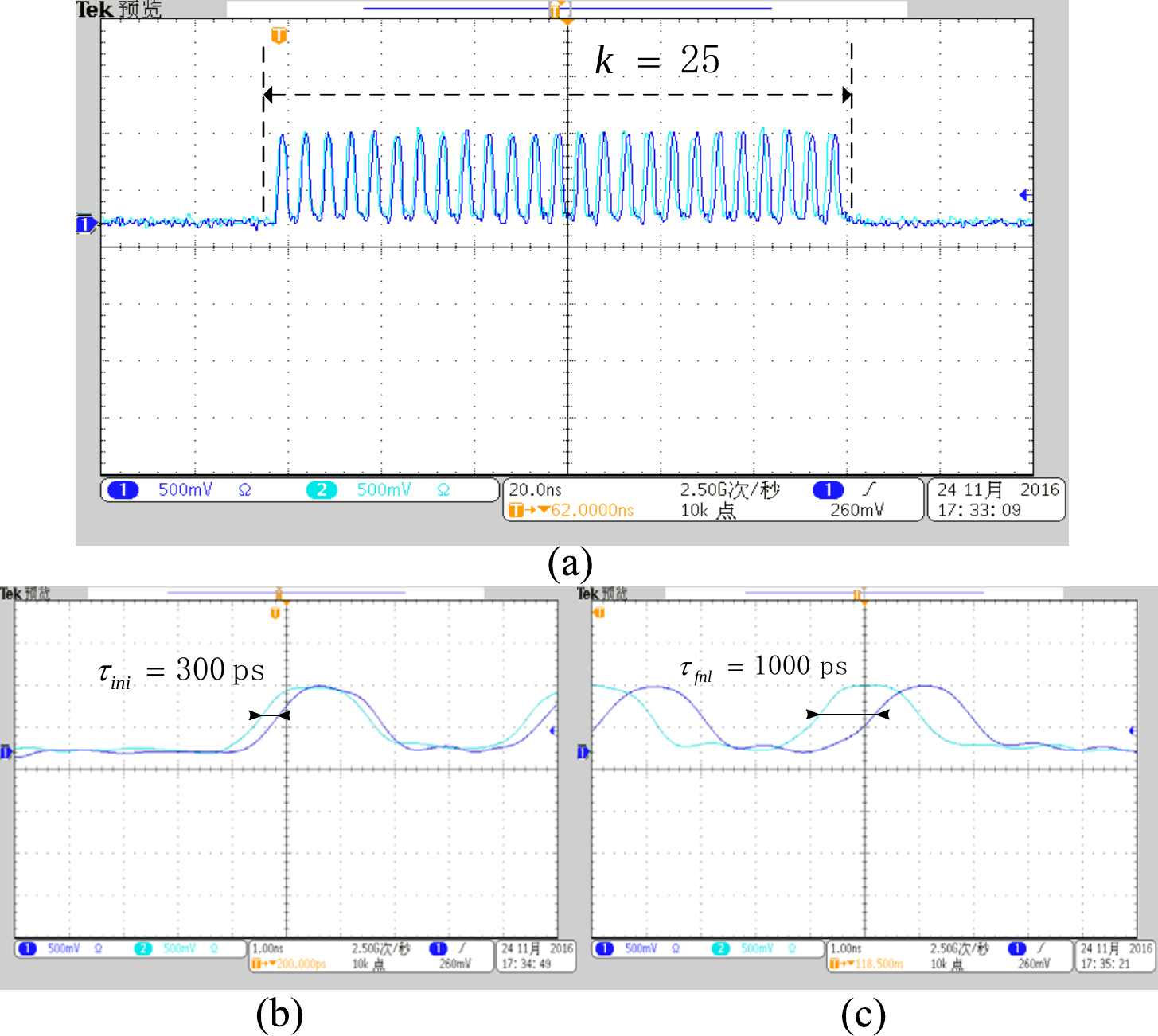}

\caption{Oscillation waveforms for the DU number combination set $(i,\thinspace j)$.}
\end{figure}

According to our design experience, $16\times16$ design space generating
256 possible DU number combination sets is large enough to construct
our 32-channels TDC. No failure happens during the whole design process.
As an example, we summarize the recorded period difference values
with the $16\times16$ design space for the TDC channel No.1 in Table
II. 

\begin{table}[tbh]
\caption{Recorded period difference values for the TDC channel No.1 in our
design }

\centering%
\begin{tabular}{|c|c||c|c|}
\hline 
$i$ (fixing $j=32$) & $\varDelta\tau_{i,32}$ (ps) & $j$ (fixing $i=32$) & $\varDelta\tau_{32,j}$ (ps)\tabularnewline
\hline 
\hline 
32 & -133 & 32 & -133\tabularnewline
\hline 
31 & -175 & 31 & -131\tabularnewline
\hline 
30 & -63 & 30 & -168\tabularnewline
\hline 
29 & -96 & 29 & -256\tabularnewline
\hline 
28 & 70 & 28 & -230\tabularnewline
\hline 
27 & 45 & 27 & -371\tabularnewline
\hline 
26 & 88 & 26 & -344\tabularnewline
\hline 
25 & 62 & 25 & -383\tabularnewline
\hline 
24 & 131 & 24 & -333\tabularnewline
\hline 
23 & 145 & 23 & -400\tabularnewline
\hline 
22 & 130 & 22 & -286\tabularnewline
\hline 
21 & 125 & 21 & -433\tabularnewline
\hline 
20 & 190 & 20 & -400\tabularnewline
\hline 
19 & 450 & 19 & -406\tabularnewline
\hline 
18 & 135 & 18 & -362\tabularnewline
\hline 
17 & 90 & 17 & -400\tabularnewline
\hline 
\end{tabular}
\end{table}

Since a target resolution range of 25 \textasciitilde{} 35 ps is chosen
in our design, by exploiting Table II, we can easily conclude the
satisfying DU number combination sets by applying equation (\ref{eq:RDR}).
For example, the combination set $(i,\thinspace j)$=(25, 30) gives
$\varDelta\tau_{25,\thinspace30}=-168+62-(-133)=27$ ps which demonstrates
itself a valid DU combination candidate. It should be noticed that
the PDR method just provides an estimation of the resolution whose
accurate value should be obtained from the code density tests as performed
in Section III.

\section{Test Results}

This paper built a 32-channels TDC prototype on a single EP3SE110F1152I3
Stratix \mbox{III} device from Altera using a self-designed test board.
The coarse counter is set as 9 bits width running at 600 MHz clock
rate (corresponding to 1667 ps period). The fine time counter is set
as 7 bits width. There are totally 16 bits to represent a timestamp.
Resource report after compilation shows the LUT occupation percentage
is 319/85200 (0.4\%) and the register occupation percentage is 104/85200
(0.2\%) per TDC channel. So only about 13\% LUTs and 7\% registers
of the FPGA chip are cost for the 32-channels TDC design. 

\begin{figure}[tbh]
\centering\includegraphics[width=0.85\columnwidth]{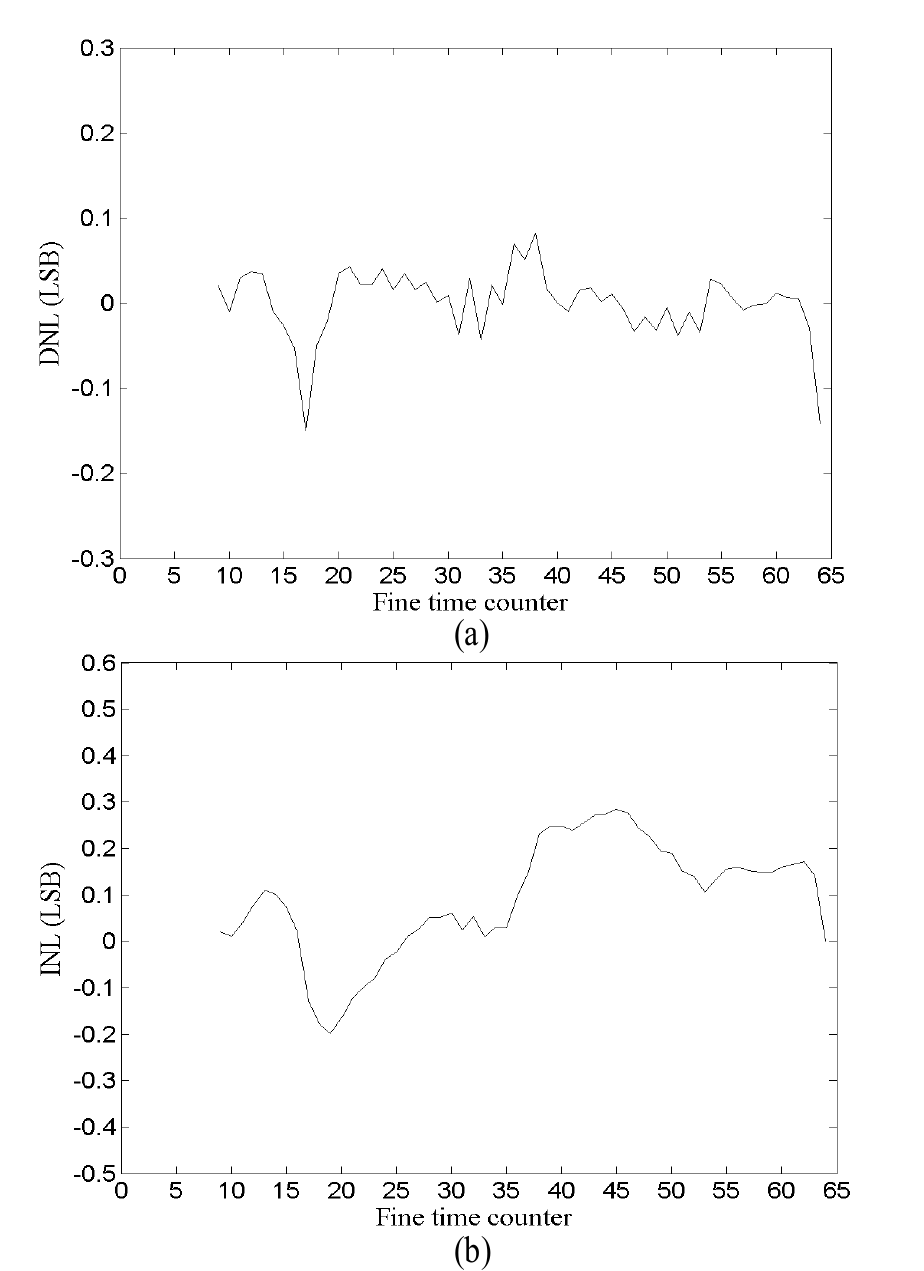}

\caption{The DNL (a) and INL (b) of the TDC channel No.1.}
\end{figure}

\begin{figure}[tbh]
\centering\includegraphics[width=1\columnwidth]{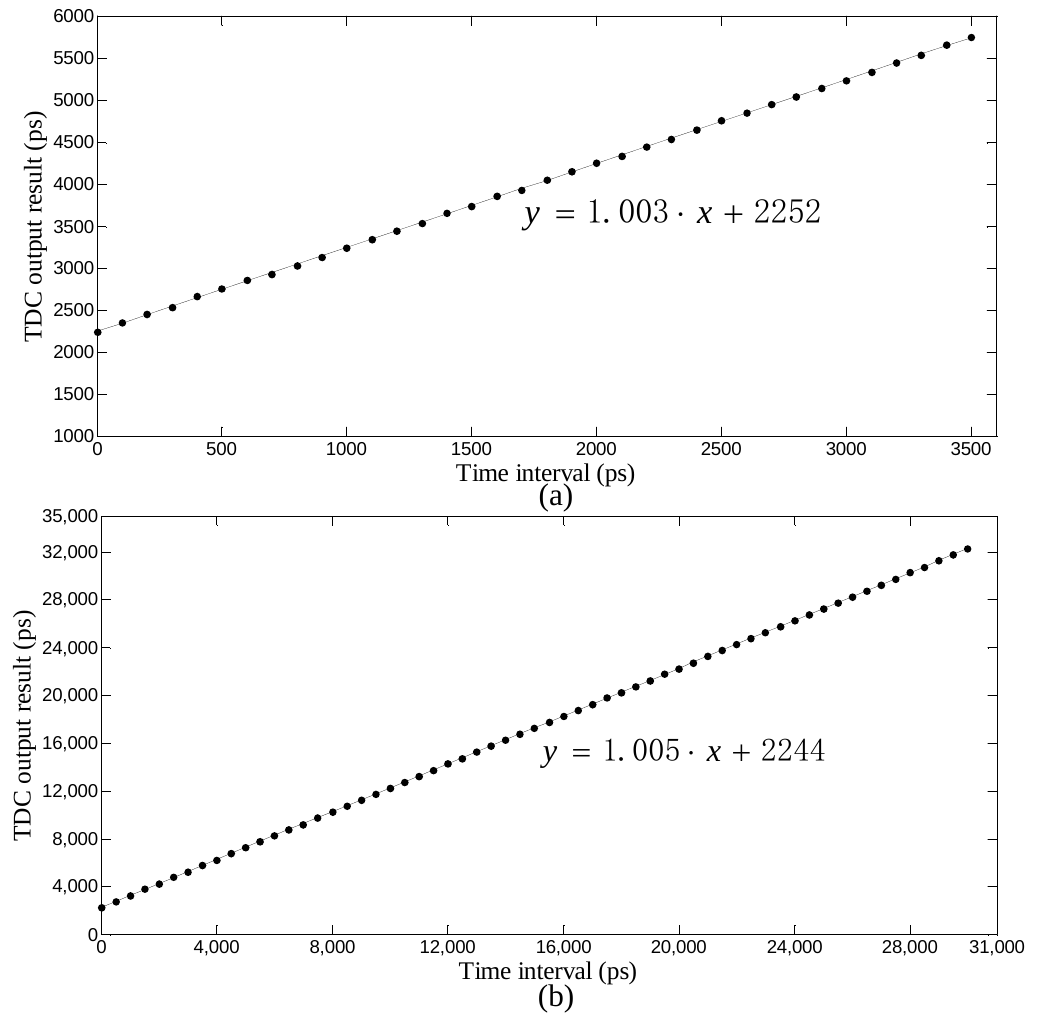}

\caption{Transfer curves. (a) using a step size of 100 ps and a dynamic range
of 3.5 ns; (b) using a step size of 500 ps and a dynamic range of
30 ns.}
\end{figure}

\begin{figure}[tbh]
\centering\includegraphics[width=0.9\columnwidth]{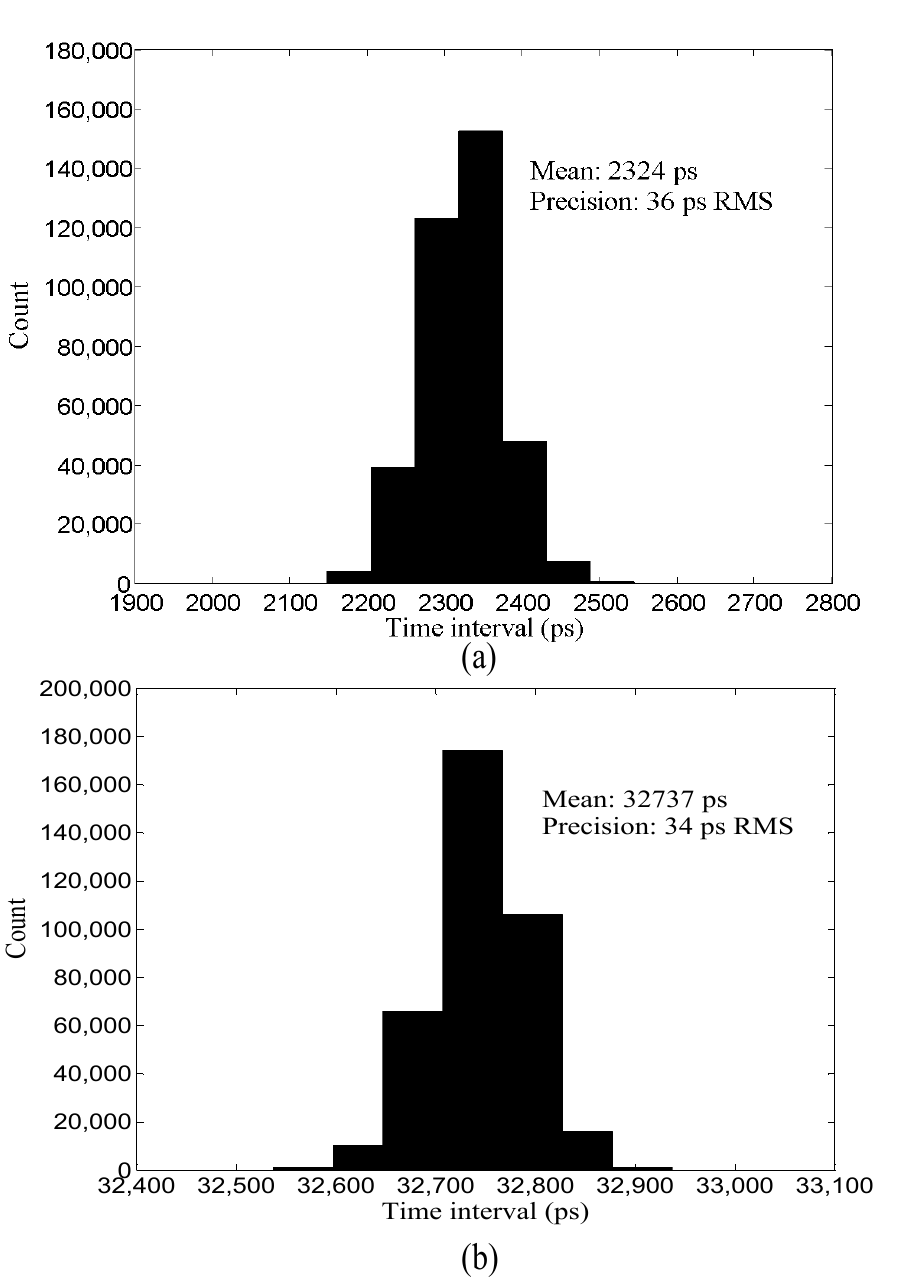}

\caption{Histogram of the time interval results between the TDC channels No.2
and No.1.}
\end{figure}

\begin{figure}[tbh]
\centering\includegraphics[width=0.95\columnwidth]{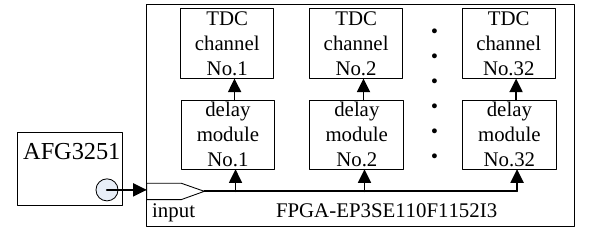}

\caption{Test configuration for the proposed 32-channels TDC.}
\end{figure}

During tests, all recorded timestamps were transferred to PC via the
USB 2.0 bus for further analysis. Code density tests were applied
to test the performance of the DNL and INL. Furthermore, precision
RMS test was also performed via two TDC channels by feeding two hit
signals having a fixed delay value. To reduce any possible statistical
error of counting, the test sample size was set to be one million.
All the mentioned tests were conducted using nominal supply voltages
and at an ambient temperature of around 20\textdegree C.

\subsection{Specific performance characterization of TDC channel No.1}

To apply the code density test, an arbitrary function generator AFG3251
was used to generated pulsed signals with repetition frequency of
500.1 kHz. The generator ran under an uncorrelated clock with the
TDC to guarantee the correctness of the code density tests. The pulsed
signals were introduced into the FPGA chip acting as hit signals.
The tested fine timestamps for TDC channel No.1 lie in range of (9
\textasciitilde{} 64), so the $\textrm{LSB}=\frac{1667\thinspace\textrm{ps}}{64-9}=30.3$
ps. The obtained diagrams of the DNL lying in the range of (-0.15
LSB \textasciitilde{} 0.82 LSB) and the INL lying in the range of
(-0.21 LSB \textasciitilde{} 0.28 LSB) are depicted in Fig.6.

To test large time interval results and evaluate the precision RMS,
TDC channel No.2 was included. The AFG3251 was used to generate two
correlated hit signals with a programmed delay value ranging in (0
ps \textasciitilde{} 30000 ps). The two hit signals were fed to TDC
channels No.1 and No.2 respectively by using two co-axial cable with
equal length. The TDC output results were obtained by subtracting
the time results of channel No.1 from those of channel No.2. Before
test, a 740Zi Lecroy digital oscilloscope working at the Random Interleaved
Sampling Mode (RIS) providing 200 Gs/s equivalent sampling rate was
used to determine the signal jitter introduced by the AFG3251 which
turned out to be less than 8 ps. That value has small influence to
our final results. The transfer curves of the TDC are depicted in
Fig.7 of which Fig.7(a) uses a step size of 500 ps and a dynamic range
of 30 ns while Fig.7(b) uses a step size of 100 ps and a dynamic range
of 3.5 ns. The fitted linear curve has a slope very close to 1 which
demonstrates that the TDC has very good linearity performance. The
offset in the figure is mainly caused by the delay path difference
of the two TDC channels from the IO element to the TDC module on the
FPGA chip. During the transfer curve test process, the precision RMS
values at each time interval point are calculated simultaneously which
turn out to lie in the range of (32 ps \textasciitilde{} 40 ps). As
an example, the histograms of the time interval results with values
of 2324 ps and 32737 ps are depicted in Fig.8. 

\begin{table*}[tbh]
\caption{Tested performance of each of the 32-channels TDC}

\begin{tabular}{|>{\centering}p{0.1\columnwidth}|c|c|>{\centering}p{0.2\columnwidth}|c|c|>{\centering}p{0.35\columnwidth}|>{\centering}p{0.35\columnwidth}|}
\hline 
channel number & $(n_{0},\thinspace n_{m})$ & LSB (ps) & equivalent 

bin width (ps) & DNL (\% LSB) & INL (\% LSB) & mean delay value (ps)

relative to channel No.1 & precision RMS (ps)

relative to channel No.1\tabularnewline
\hline 
\hline 
1 & (9, 64) & 30.3 & 30.5 & -15\textasciitilde{}8 & -20\textasciitilde{}28 & N/A & N/A\tabularnewline
\hline 
2 & (13, 60) & 35.5 & 35.7 & -7\textasciitilde{}9 & -3\textasciitilde{}51 & 5182 & 34\tabularnewline
\hline 
3 & (12, 76) & 26.0 & 26.3 & -22\textasciitilde{}24 & -38\textasciitilde{}40 & 5689 & 38\tabularnewline
\hline 
4 & (12, 64) & 32.1 & 32.3 & -10\textasciitilde{}11 & -26\textasciitilde{}28 & 6891 & 36\tabularnewline
\hline 
5 & (8, 64) & 29.7 & 30.0 & -15\textasciitilde{}12 & -72\textasciitilde{}5 & 7231 & 34\tabularnewline
\hline 
6 & (9, 60) & 32.7 & 33.1 & -32\textasciitilde{}27 & -51\textasciitilde{}36 & 7983 & 35\tabularnewline
\hline 
7 & (9, 58) & 34.0 & 34.1 & -8\textasciitilde{}8 & -2\textasciitilde{}47 & 6420 & 33\tabularnewline
\hline 
8 & (8, 59) & 32.7 & 32.9 & -11\textasciitilde{}13 & -28\textasciitilde{}30 & 8821 & 32\tabularnewline
\hline 
9 & (8, 68) & 27.8 & 30.0 & -9\textasciitilde{}9 & -55\textasciitilde{}16 & 8136 & 38\tabularnewline
\hline 
10 & (6, 68) & 26.9 & 27.2 & -22\textasciitilde{}16 & -28\textasciitilde{}33 & 9247 & 36\tabularnewline
\hline 
11 & (6, 57) & 32.7 & 32.9 & -10\textasciitilde{}10 & -48\textasciitilde{}11 & 9835 & 32\tabularnewline
\hline 
12 & (10, 59) & 34.0 & 34.2 & -13\textasciitilde{}15 & -58\textasciitilde{}7 & 10125 & 34\tabularnewline
\hline 
13 & (12, 65) & 31.5 & 31.6 & -9\textasciitilde{}9 & -51\textasciitilde{}22 & 11678 & 36\tabularnewline
\hline 
14 & (15, 73) & 28.7 & 29.0 & -19\textasciitilde{}22 & -52\textasciitilde{}12 & 12872 & 39\tabularnewline
\hline 
15 & (10, 60) & 33.3 & 33.6 & -18\textasciitilde{}22 & -46\textasciitilde{}20 & 14883 & 33\tabularnewline
\hline 
16 & (4, 58) & 30.9 & 31.2 & -33\textasciitilde{}16 & -7\textasciitilde{}65 & 15824 & 33\tabularnewline
\hline 
17 & (5, 56) & 32.7 & 32.9 & -13\textasciitilde{}11 & -62\textasciitilde{}12 & 13764 & 36\tabularnewline
\hline 
18 & (4, 59) & 30.3 & 30.5 & -18\textasciitilde{}12 & -10\textasciitilde{}69 & 12748 & 37\tabularnewline
\hline 
19 & (9, 58) & 34.0 & 34.3 & -21\textasciitilde{}14 & -30\textasciitilde{}51 & 14782 & 33\tabularnewline
\hline 
20 & (11, 60) & 34.0 & 34.3 & -15\textasciitilde{}16 & -68\textasciitilde{}14 & 15824 & 34\tabularnewline
\hline 
21 & (8, 76) & 24.5 & 24.7 & -20\textasciitilde{}23 & -59\textasciitilde{}25 & 16732 & 35\tabularnewline
\hline 
22 & (7, 52) & 37.0 & 37.2 & -21\textasciitilde{}36 & -12\textasciitilde{}41 & 15320 & 35\tabularnewline
\hline 
23 & (8, 80) & 23.2 & 23.3 & -13\textasciitilde{}18 & -21\textasciitilde{}38 & 16329 & 37\tabularnewline
\hline 
24 & (6, 70) & 26.0 & 26.2 & -13\textasciitilde{}21 & -17\textasciitilde{}33 & 15897 & 38\tabularnewline
\hline 
25 & (8, 57) & 34.0 & 34.2 & -21\textasciitilde{}16 & -51\textasciitilde{}10 & 17231 & 35\tabularnewline
\hline 
26 & (10, 58) & 34.7 & 34.9 & -32\textasciitilde{}12 & -44\textasciitilde{}24 & 17467 & 33\tabularnewline
\hline 
27 & (8, 58) & 33.3 & 33.4 & -21\textasciitilde{}30 & -22\textasciitilde{}18 & 18237 & 32\tabularnewline
\hline 
28 & (10, 66) & 29.8 & 29.9 & -19\textasciitilde{}12 & -40\textasciitilde{}38 & 15983 & 36\tabularnewline
\hline 
29 & (9, 61) & 32.1 & 32.3 & -14\textasciitilde{}15 & -54\textasciitilde{}31 & 18476 & 37\tabularnewline
\hline 
30 & (5, 58) & 31.5 & 31.7 & -20\textasciitilde{}17 & -18\textasciitilde{}36 & 19238 & 38\tabularnewline
\hline 
31 & (11, 75) & 26.0 & 26.2 & -13\textasciitilde{}21 & -26\textasciitilde{}33 & 19782 & 36\tabularnewline
\hline 
32 & (7, 55) & 34.7 & 34.8 & -13\textasciitilde{}17 & -12\textasciitilde{}17 & 18472 & 32\tabularnewline
\hline 
\end{tabular}

\ 

N/A=not applicable
\end{table*}

\subsection{Performance summarization of all the 32 TDC channels}

In this section, a specific test configuration depicted in Fig.9 was
applied to help simplify the test process. The AFG3251 is used to
generate hit signals with repetition frequency of 500.1 kHz. We set
a delay module independently for each of the 32 TDC channels which
is composed of cascaded NOT gates with even number (the gates number
is randomly set in the range of 40 \textasciitilde{} 100). This configuration
is very useful to evaluate the precision RMS under large time interval
tests such as 4 \textasciitilde{} 20 ns. 

By analyzing the distribution of the time timestamps for each of the
32 TDC channels, all important performance parameters including fine
time counter range, resolution, equivalent bin width, DNL, INL and
precision RMS can be obtained. The detailed parameters results are
listed in Table III. In this table, the resolution is calculated as
$\textrm{LSB}=\frac{1667\thinspace\textrm{ps}}{n_{m}-n_{0}}$. The
term equivalent bin width $w_{eq}$ can take effects of the various
bin widths into account \cite{Jwu2014Uneven}. It is calculated as
$w_{eq}=\sqrt{\sum_{i}(\frac{w_{i}^{3}}{W})}$ with $W=\sum_{i}w_{i}$,
where $w_{i}$ represents the bin width for the $i$-th bin number.
All the $w_{i}$ values are obtained by the code density tests. 

From Table III, we conclude that the obtained resolutions and equivalent
bin widths all lie in the range of (23.2 ps \textasciitilde{} 37.2
ps), and the fact that they are very close to each other reflects
that the TDC has good linearity performance \cite{QiShen2015}. The
obtained DNL results generally lie in the range of (-0.4 LSB \textasciitilde{}
0.4 LSB) with a maximal amplitude of 0.59 LSB (channel number 6) and
the obtained INL results generally lie in the range of (-0.7 LSB \textasciitilde{}
0.7 LSB) with a maximal amplitude of 0.87 LSB (channel number 6).
The obtained linearity is not as good as that reported in \cite{Cui2017}.
One reason is that the physical location of a TDC channel on the FPGA
chip is found to influence the linearity error significantly. However,
we did not optimize the physical locations in this design since it
would be considerably time consuming and not necessarily required
in most application cases. All the implementation regions were automatically
generated by the compilation software. If the designers want to obtain
TDC channels with very small linearity error, manually assigning the
implementation regions and comparing their performance are recommended.
Another reason is that multi-channels may influence each other during
operation and deteriorate the linearity error. Manually and properly
assign the implementation regions may help improve the linearity performance.
Even so the linearity performance is still relatively better than
that in the TDL based method utilizing carry chains which usually
owns a maximal amplitude of 2 \textasciitilde{} 4 LSBs. 

Large time interval results are obtained by subtracting the time results
calculated from TDC channel No.1 from those of the TDC channels No.2
\textasciitilde{} 32 respectively. The precision RMS is calculated
from the corresponding time interval results for each of the TDC channels.
From Table III, it can be seen that all of the precision RMS results
lie in the range of (32 ps \textasciitilde{} 39 ps).

Finally, the dead time of the realized TDC channels is mainly determined
by the oscillation period of the Vernier delay line and the maximal
oscillation numbers. The oscillation period (from Fig.5) is about
5 ns and the maximal oscillation number is 80 (from Table III, channel
No.23) leading to the dead time of $5\times80=400$ ns.

\section{Discussion}

Carry chains are usually organized in TDL style which is the mainstream
realization method for FPGA-based TDCs. This method provides low implementation
complexity since the carry chain based TDL can be automatically synthesized
by software compiler without any manual intervention. However, a plain
TDC constructed by the TDL method usually suffers from large DNL and
INL. Fortunately, by applying some well developed optimization techniques,
such as the wave union \cite{JWu2008} or multi-chains averaging technique
\cite{QiShen2015} to improve the equivalent resolution and the bin-by-bin
calibration technique \cite{Wu2010} to improve the INL, this TDC
method is very promising for practical use. 

This paper emphasizes the Vernier method by organizing the carry chains
in RO style. This method has demonstrated itself very competitive
in terms of resource cost, DNL and INL when compared with the TDL
method for a plain TDC design. The shortcomings are that the realized
resolution is not as high as that in the TDL method so far, the dead
time is relatively longer, and manual intervention to adjust the RO
period difference is needed during design process. However, similar
optimization techniques such as the multi-chains averaging and bin-by-bin
calibration can also be applied to this kind of TDC to further improve
its performance. Most importantly, applying the multi-chains averaging
technique is very valuable to suppress the large precision RMS and
further exploits the resolution capability of such TDCs down to 10
ps level. Some performance comparisons between this work and some
other recent FPGA-based works are summarized in Table IV.

\begin{table*}[tbh]
\caption{Performance comparisons between this work and some other recent FPGA-based
works}

\begin{tabular}{|c|c|c|>{\centering}p{0.2\columnwidth}|>{\centering}p{0.15\columnwidth}|>{\centering}p{0.2\columnwidth}|>{\centering}p{0.2\columnwidth}|>{\centering}p{0.2\columnwidth}|>{\centering}p{0.15\columnwidth}|>{\centering}p{0.1\columnwidth}|}
\hline 
ref. & chip & method & resolution 

(ps) & precision 

RMS (ps) & DNL

(LSB) & INL

(LSB) & dead time

(ns) & costed 

registers & costed LUTs\tabularnewline
\hline 
\hline 
\cite{LZhao2013} & Virtex-5 & TDL & 30 & 15 & -1 \textasciitilde{} 3 & -4 \textasciitilde{} 4  & 30 & 571 & 1064\tabularnewline
\hline 
\cite{YonggangW2016-2} & UltraScale & TDL & 4.5 & 3.9 & N/S & N/S & 4 & N/S & N/S\tabularnewline
\hline 
\cite{7057685} & Virtex-6 & TDL & 10 & 12.8 & -1 \textasciitilde{} 1.91 & -2.2 \textasciitilde{} 3.93  & N/S & N/S & N/S\tabularnewline
\hline 
\cite{WPan2014} & Cyclone II & TDL & 21.8 & 28.8 & N/S & N/S & N/S & 23494 & 28085\tabularnewline
\hline 
\cite{MFishburn2013} & Virtex-6 & TDL & 10 & 19.6 & -1 \textasciitilde{} 1.5  & -2.25 \textasciitilde{} 1.61 & 3.3 & N/S & N/S\tabularnewline
\hline 
\cite{SzpletKlepacki2010} & Spartan-3 & pulse shrinking & 42 & 56 & -0.98 \textasciitilde{} 0.6 & -4.17 \textasciitilde{} 3.6 & 710 & N/S & N/S\tabularnewline
\hline 
this work & Stratix III & Vernier & 23 \textasciitilde{} 37 & 32 \textasciitilde{} 39 & -0.4 \textasciitilde{} 0.4 & -0.7 \textasciitilde{} 0.7 & 400 & 319 & 104\tabularnewline
\hline 
\end{tabular}

\,\,\,\,\,N/S=not specified
\end{table*}

\section{Conclusions}

Our recently proposed RO-based TDCs by organizing the carry chains
in the Vernier loop style are a promising option for the TDC designers
mainly due to its remarkably low linearity error and low resource
cost. However, implementation complexity problem is posed since this
design calls for manual intervention to the initial fitting results
when moving to multi-channels TDC designs. To combat that problem,
this paper elaborates the key points to construct multi-channels TDCs
to achieve high performance while keeping the least design complexity:
one for the clock extraction module and one for the fine time interpolator
module. Furthermore the PDR method is proposed to search the potential
DU number combination sets for a targeted resolution which costs at
most 31 trials in our example design. The PDR method greatly reduces
the implementation complexity during the fine time interpolator construction
process. This paper built a 32-channels TDC on a Altera Stratix III
FPGA and demonstrates good performance. This paper greatly eases the
designing difficulty of the carry chain RO-based TDCs and can significantly
propel their development in practical use.


\end{document}